\documentclass[conference]{IEEEtran}
\IEEEoverridecommandlockouts

\usepackage{amsmath,amsfonts}
\usepackage{algorithmic}
\usepackage{algorithm}
\usepackage{array}
\usepackage[caption=false,font=normalsize,labelfont=sf,textfont=sf]{subfig}
\usepackage{textcomp}
\usepackage{stfloats}
\usepackage{url}
\usepackage{verbatim}
\usepackage{graphicx}
\usepackage{romannum}
\usepackage{cite}
\usepackage{makecell}
\usepackage{multirow}
\usepackage{xcolor}
\def\BibTeX{{\rm B\kern-.05em{\sc i\kern-.025em b}\kern-.08em
    T\kern-.1667em\lower.7ex\hbox{E}\kern-.125emX}}

\title{Deployment-Efficient Short-Term Load Forecasting in AI Data Centers via Sequence-to-Point Knowledge Distillation}

\author{
\IEEEauthorblockN{Lei Wang, Jiahao Chen}
\IEEEauthorblockA{\textit{School of Electrical and Computer Engineering} \\
\textit{Oklahoma State University} \\
Stillwater, OK, U.S. \\
\{leiwang; jiahao.chen\}@okstate.edu}\\
\IEEEauthorblockN{Fanping Sui}
\IEEEauthorblockA{\textit{Department of Mechanical Engineering} \\
\textit{University of California Berkeley} \\
Berkeley, CA, U.S. \\
fpsui@berkeley.edu}\\

\and

\IEEEauthorblockN{Ying Zhang*}
\IEEEauthorblockA{\textit{School of Electrical and Computer Engineering} \\
\textit{Oklahoma State University} \\
Stillwater, OK, U.S. \\
y.zhang@okstate.edu}\\

\IEEEauthorblockN{Di Shi}
\IEEEauthorblockA{\textit{School of Electrical and Computer Engineering} \\
\textit{New Mexico State University} \\
Las Cruces, NM, U.S.  \\
dshi@nmsu.edu}
\thanks{This research was supported by the U.S. National Science Foundation through award number 2418359 and the Hamm Institute of American Energy.

*Corresponding author: Ying Zhang}
}

\begin{document}
\maketitle

\begin{abstract}
Accurately forecasting the bursty and non-stationary power demand of AI data centers has become increasingly important, as abrupt workload-driven variations at the GPU-node level can affect real-time operational efficiency, power management, and grid-data center coordination. However, high-capacity forecasting models are often difficult to deploy at scale because of their memory and latency requirements, while lightweight predictors may fail to capture short-horizon temporal dynamics. To address this accuracy-deployment tradeoff, this paper proposes a deployment-efficient knowledge distillation framework for short-term load forecasting in AI data centers. The proposed framework first trains a high-capacity sequence teacher model for multi-step load trajectory prediction, where residual learning is used to improve robustness under non-stationary operating conditions. A lightweight point-wise student model is then developed for low-latency rolling inference using a compact neural network architecture. To transfer temporal knowledge from the teacher to the student, a sequence-to-point distillation strategy is introduced by aligning near-term predictive behavior and temporally pooled representations. Case studies on the MIT Supercloud dataset demonstrate that the proposed student model improves forecasting accuracy over recent deep learning baselines while reducing the deployment footprint by over 10× in parameter memory and model size.

\end{abstract}


\section{Introduction}

\IEEEPARstart{W}{ith} the rapid proliferation of AI training, data centers are emerging as highly dynamic electric loads with strong non-stationarity and abrupt ramping behavior \cite{Seshmasetti2025, Chen2025}. Unlike conventional commercial loads, AI data center demand is driven by fast-changing and often hidden workload dynamics. At minute-level or shorter time scales, job switching, concurrency variations, and hardware power management can trigger abrupt power jumps and persistent fluctuations, leading to bursty load patterns that are difficult to forecast \cite{Takci2025}. These characteristics create new requirements for both node-level job management and cluster-level grid-data center coordination. In particular, accurate short-horizon load forecasting is essential for real-time workload scheduling, power smoothing, and demand response \cite{Chen2020}. However, deploying high-fidelity forecasting models at the GPU-node level remains challenging because online inference must satisfy tight memory and latency constraints as AI workloads continue to scale.

Conventional electric load forecasting methods often rely on predefined temporal structures, such as autoregressive models and exponential smoothing \cite{Suganthi2012, Jadon2021}. While these approaches are typically computationally efficient, their performance often depends on assumptions such as local linearity, stationarity, and fixed noise distributions. Such assumptions are frequently violated in AI data center workloads, where job switching, concurrency changes, and hardware power management can induce bursty transients and regime shifts \cite{Chen2025}. To better handle nonlinear and non-stationary behavior, data-driven forecasting methods have been increasingly adopted, including tree-based machine-learning models \cite{Mollasalehi2025}, recurrent neural networks such as LSTM and GRU \cite{Yunita2025}, and temporal convolution-based models such as temporal convolutional networks (TCNs) \cite{Ghimire2025}. More recently, these learning-based approaches have also begun to be explored for AI data center load forecasting.

A recent study \cite{mughees2025short} investigated the use of long short-term memory (LSTM) and gated recurrent unit (GRU) networks for forecasting aggregated GPU-node power consumption. However, reported results show that these models still struggle to capture sudden ramps, dips, and peaks. More broadly, many existing estimates of AI electricity demand are based on aggregate GPU counts or annual energy use \cite{mughees2025short,mural2026ai}, which limits node-level awareness and reduces opportunities for flexible job management and power smoothing. At the individual computing-node level, job switching and concurrency variations can induce abrupt ramps and regime shifts, making short-horizon load dynamics highly non-stationary and substantially more challenging than cluster-level forecasting \cite{Chen2025}. Although high-capacity models such as Transformers are well suited to capturing such dynamics, their online execution can be computationally expensive and may impose additional operational burdens as the number of computing nodes continues to grow \cite{Jadon2021}. In contrast, lightweight point predictors are attractive for deployment but often cannot adequately capture trajectory-level temporal dynamics, especially during bursty transients. As a result, balancing forecasting accuracy and deployment efficiency remains a key challenge for short-term load forecasting in AI data centers \cite{Ramani2025}.

To address this challenge, this paper proposes a deployment-efficient short-term load forecasting framework for AI data centers that transfers sequence-level temporal knowledge learned offline into a compact predictor suitable for rolling real-time inference. The proposed method follows an offline-to-online teacher--student design at the GPU-node level. A high-capacity sequence teacher model is first trained offline to learn short-horizon load trajectories under bursty and non-stationary operating conditions. A lightweight point-wise student model is then distilled for low-latency online inference using a compact neural network architecture suitable for deployment at scale. To enable sequence-to-point knowledge transfer, the proposed distillation strategy aligns both near-term predictive behavior and temporally pooled latent representations between the teacher and student models. The main contributions of this paper are summarized as follows:
\begin{itemize}
\item A sequence-to-point teacher--student knowledge distillation framework is developed to transfer trajectory-level temporal knowledge from an offline multi-step sequence model to a lightweight point-wise predictor for rolling real-time inference.

\item A residual-learning-based forecasting design is introduced to improve robustness under bursty and non-stationary AI data center operating conditions.

\item Case studies on the MIT Supercloud dataset demonstrate that the proposed student model achieves a favorable accuracy--efficiency tradeoff, outperforming recurrent baselines while substantially reducing parameter memory, on-disk model size, and CPU inference latency. 
\end{itemize}

\section{Problem Formulation}

Let $P_t$ denote the power consumption of a GPU node in an AI data center at time step $t$. 
Given a historical observation window of length $L$, the past load sequence is denoted by
$\mathbf{P}_{t-L+1:t}=\{P_{t-L+1},\ldots,P_t\}$. In addition, a multivariate exogenous feature vector $\mathbf{x}_t$ is observed at time step $t$, which is constructed from telemetry variables and workload-related indicators such as utilization levels, device temperatures, job statistics, and time encodings. The aligned historical feature sequence is denoted by $\mathbf{x}_{t-L+1:t}$.

This paper focuses on short-term GPU-node load forecasting in AI data centers to support real-time job management. The forecasting model is therefore expected to capture bursty and non-stationary load variations driven by workload dynamics. The short-term forecasting task is formulated over a finite prediction horizon of $H$ time steps:
\begin{equation}
\hat{\mathbf{P}}_{t+1:t+H}
= f\!\left(\mathbf{P}_{t-L+1:t},\, \mathbf{x}_{t-L+1:t}\right)
\label{eq:multi_step_forecast}
\end{equation}
where $\hat{\mathbf{P}}_{t+1:t+H}=\{\hat{P}_{t+1},\ldots,\hat{P}_{t+H}\}$ denotes the predicted load trajectory over the next $H$ intervals. This multi-step formulation enables the forecasting model to learn trajectory-level temporal patterns, including local ramps, burst onsets, and short-horizon regime transitions in AI data center loads.

In practical short-term forecasting, forecasts are often updated in a rolling manner. 
A general one-step forecasting formulation can be written as
\begin{equation}
\hat{P}_{t+1}
= g\!\left(\mathbf{P}_{t-L+1:t},\, \mathbf{x}_{t-L+1:t}\right)
\label{eq:one_step_forecast}
\end{equation}
where $g(\cdot)$ denotes a generic sequence-input one-step forecasting function based on the historical load and feature sequences. 
By recursively applying this one-step formulation, a short future trajectory can be generated in a rolling manner.

The above formulations highlight a trade-off between temporal modeling capability and deployment efficiency. 
Multi-step sequence forecasting provides richer trajectory-level supervision and is well-suited to learning complex short-horizon workload dynamics. 
However, directly deploying sequence-input forecasting models at each GPU node may introduce non-negligible memory and latency costs, particularly when forecasts are updated repeatedly in a rolling manner. 
In contrast, a point-wise one-step predictor using only $(P_t,\mathbf{x}_t)$ is more suitable for low-latency online inference, but it may lose trajectory-level temporal information under bursty and non-stationary operating conditions.

To address this challenge, this paper develops a knowledge-distilled short-term forecasting approach in which an offline multi-step sequence teacher learns trajectory-level temporal dynamics, while the deployed student is constrained to a lightweight point-wise input structure. 
The proposed sequence-to-point distillation strategy transfers the temporal richness learned by the sequence teacher into the point-wise student predictor for low-latency rolling inference.

\begin{figure}[!t]
\vspace{5pt}
\includegraphics[width=0.47\textwidth]{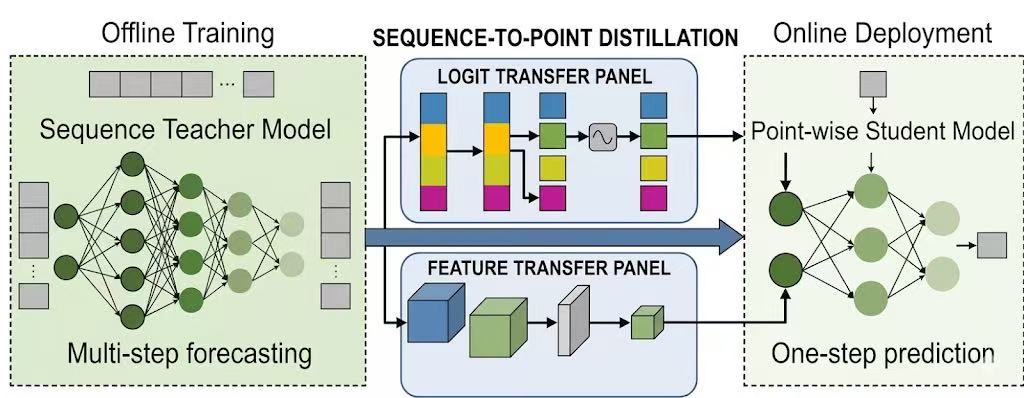}
\caption{Schematic diagram of the proposed teacher-student knowledge distillation framework.}
\label{Fig.1}
\end{figure}
\section{Proposed Sequence-to-Point Knowledge Distillation Method}

The proposed method transfers predictive trajectory knowledge learned from an offline multi-step sequence forecasting teacher model to a one-step student predictor. As illustrated in Fig.~\ref{Fig.1}, the proposed framework adopts an offline-to-online teacher-student design to bridge the gap between sequence-level temporal learning and point-wise deployment.
A high-capacity sequence teacher model is first trained offline to capture short-horizon temporal dynamics from historical load and telemetry sequences, and a lightweight point-wise student model is then distilled for low-latency rolling deployment.
Hence, the proposed student predictor remains compact in structure while still inheriting part of the teacher model's trajectory-level temporal knowledge.

\subsection{Residual Learning-Based Teacher and Student Models}

AI data center loads often exhibit regime shifts and abrupt level variations induced by changing workloads.
Under such conditions, directly regressing the absolute future load may be less robust to short-term non-stationary variations \cite{mughees2025short}.
To improve robustness, a knowledge distillation algorithm is proposed within a residual learning framework, consisting of an attention-based sequence model as the teacher and a lightweight pointwise student model. The attention-based sequence model is selected as the teacher model because it can exploit long-range temporal dependencies and heterogeneous interactions among historical observations, without being constrained by the online latency requirements of the deployed predictor.

For each prediction horizon $h\in\{1,\ldots,H\}$, the future load is decomposed into the latest observed load and a predicted increment:
\begin{equation}
\hat{P}_{t+h} = P_t + \Delta \hat{P}_{t+h}
\label{eq:residual_formulation}
\end{equation}
where $\Delta \hat{P}_{t+h}$ denotes the predicted residual with respect to the most recent observation $P_t$.
This formulation makes the forecasting target more locally stationary and naturally aligns with rolling deployment, since each prediction is anchored to the latest available operating point.

\textbf{Sequence teacher model.}
The proposed teacher model adopts a state-of-the-art encoder-only Transformer backbone to predict the residuals and then an $H$-step  trajectory via \eqref{eq:residual_formulation}. The historical load and exogenous computing-related features are concatenated at each time step to form token vectors:
\begin{equation}
\mathbf{u}_{\tau} = [P_{\tau};\, \mathbf{x}_{\tau}] \in \mathbb{R}^{d_u},
\quad \tau = t-L+1,\ldots,t
\label{eq:token_construction}
\end{equation}
where the computing-related feature vector $\mathbf{x}_{\tau}$ includes GPU and memory utilization, temperature, and workload indicators such as job count, job switching, and GPU count.

These raw tokens are then mapped into the model space through a linear projection and positional encoding:
\begin{equation}
\mathbf{h}^{(0)}_{\tau}
= \mathbf{W}_{\mathrm{in}} \mathbf{u}_{\tau}
+ \mathbf{b}_{\mathrm{in}}
+ \mathbf{p}_{\tau}
\label{eq:input_embedding}
\end{equation}\\

\noindent where $\mathbf{h}^{(0)}_{\tau}\in\mathbb{R}^{d_m}$ denotes the initial token embedding and $\mathbf{p}_{\tau}$ provides temporal order information.

The teacher model, i.e., an encoder-only Transformer \cite{Wang2022}, adopts $N_T$ stacked encoder blocks.
Let
$\mathbf{H}^{(\ell)}=\big[\mathbf{h}^{(\ell)}_{t-L+1},\ldots,\mathbf{h}^{(\ell)}_{t}\big]^{\top}\in\mathbb{R}^{L\times d_m}$
denote the hidden states at layer $\ell$.
Each block applies multi-head self-attention followed by a position-wise feedforward network, together with residual connections and layer normalization:
\begin{equation}
\tilde{\mathbf{H}}^{(\ell)}
= \mathrm{LN}\!\left(
\mathbf{H}^{(\ell-1)} + \mathrm{MHA}\!\left(\mathbf{H}^{(\ell-1)}\right)
\right)
\label{eq:encoder_mha}
\end{equation}
\begin{equation}
\mathbf{H}^{(\ell)}
= \mathrm{LN}\!\left(
\tilde{\mathbf{H}}^{(\ell)} + \mathrm{FFN}\!\left(\tilde{\mathbf{H}}^{(\ell)}\right)
\right)
\label{eq:encoder_ffn}
\end{equation}
for $\ell=1,\ldots,N_T$, where $\mathrm{FFN}$ denotes the position-wise feedforward network
\begin{equation}
\mathrm{FFN}(\mathbf{h})
=\mathbf{W}_2\,\sigma(\mathbf{W}_1\mathbf{h}+\mathbf{b}_1)+\mathbf{b}_2
\label{eq:ffn}
\end{equation}
with nonlinearity $\sigma(\cdot)$, such as Gaussian error linear unit (GELU) or rectified linear unit (ReLU).

For completeness, the scaled dot-product attention operator is written as
\begin{equation}
\mathrm{Attn}(\mathbf{Q}, \mathbf{K}, \mathbf{V})
= \mathrm{softmax}\!\left(
\frac{\mathbf{Q}\mathbf{K}^{\top}}{\sqrt{d_k}}
\right)\mathbf{V}
\label{eq:scaled_dot_attention}
\end{equation}
and the multi-head self-attention module with $K$ heads is calculated by
\begin{equation}
\mathrm{MHA}(\mathbf{H})
= \mathrm{Concat}\!\left(
\mathrm{Attn}(\mathbf{H}\mathbf{W}^Q_k,\mathbf{H}\mathbf{W}^K_k,\mathbf{H}\mathbf{W}^V_k)
\right)_{k=1}^{K}\mathbf{W}^O
\label{eq:mha}
\end{equation}
where $\mathbf{W}^Q_k,\mathbf{W}^K_k,\mathbf{W}^V_k$, and $\mathbf{W}^O$ are learnable projection matrices.

After the stacked encoder blocks, the proposed teacher model aggregates token-wise temporal information through attention pooling:
\begin{equation}
\alpha_{\tau}
=\frac{\exp\!\left(\mathbf{w}_p^{\top}\mathbf{h}^{(N_T)}_{\tau}\right)}
{\sum_{j=t-L+1}^{t}\exp\!\left(\mathbf{w}_p^{\top}\mathbf{h}^{(N_T)}_{j}\right)}
\label{eq:attn_pool_weight}
\end{equation}
\begin{equation}
\mathbf{c}^{(T)}_{t}
=\sum_{\tau=t-L+1}^{t}\alpha_{\tau}\mathbf{h}^{(N_T)}_{\tau}\in\mathbb{R}^{d_m}
\label{eq:context_vector}
\end{equation}
where $\alpha_{\tau}$ denotes the normalized importance coefficient of the $\tau$-th historical token in attention pooling; $\mathbf{c}^{(T)}_{t}$ denotes the aggregated temporal context vector of the teacher model at time $t$, which summarizes informative historical patterns extracted by the attention-based encoder and pooling module.

\begin{figure*}
\centering
  \vspace{5pt}
\includegraphics[width=0.68\textwidth]{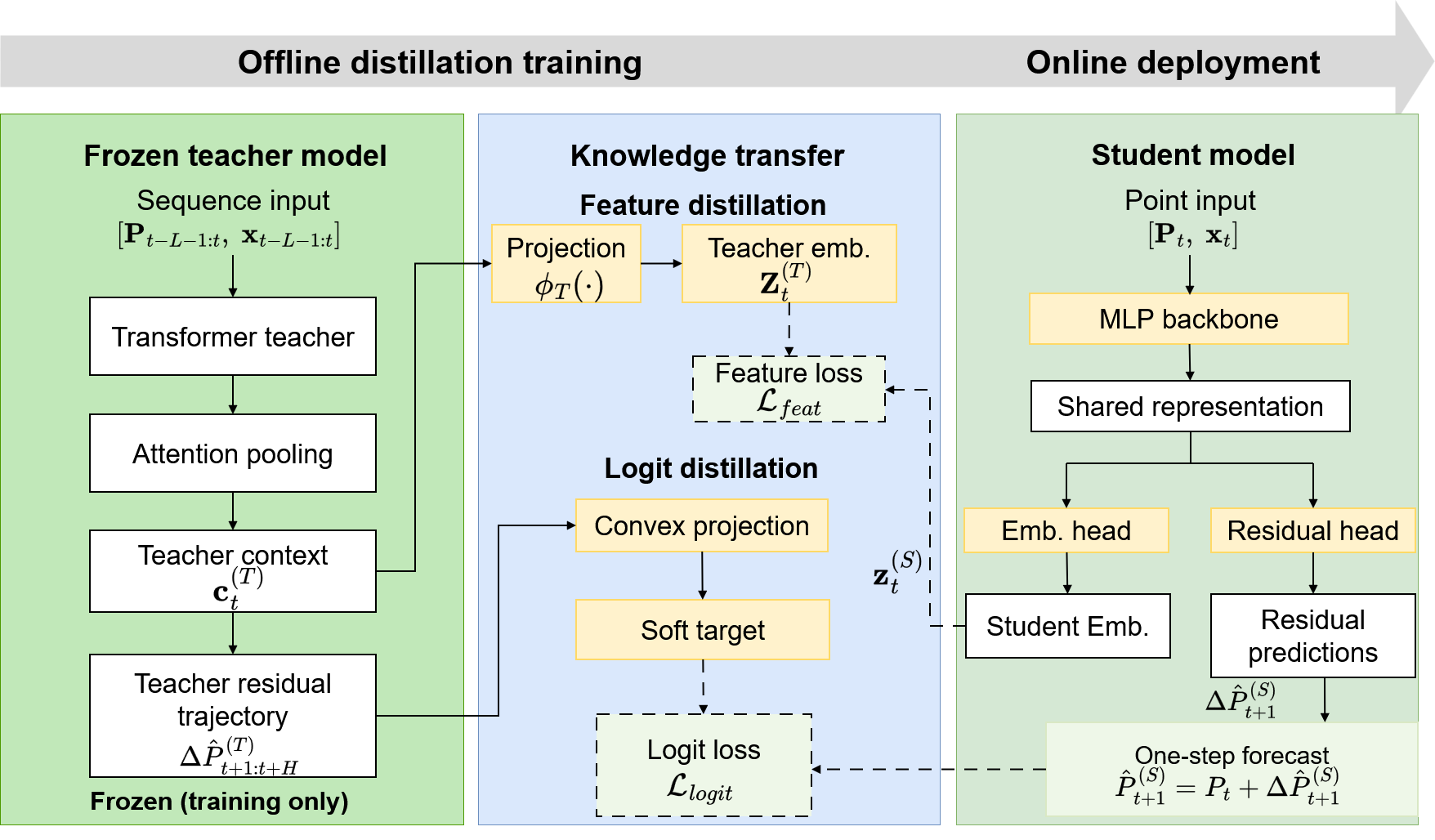}
\caption{The proposed lightweight student model distilled from the high-capacity teacher model.}
\label{Fig.2}
\end{figure*}

Compared with simply taking the last hidden state, this pooling mechanism allows the teacher to emphasize the most informative time steps for short-horizon forecasting, such as local burst onsets, workload transitions, or recent ramp patterns.
The aggregated context vector is then mapped to an $H$-step residual trajectory by a linear prediction head:
\begin{equation}
\Delta\hat{\mathbf{P}}^{(T)}_{t+1:t+H}
=\mathbf{W}_{\mathrm{out}}\mathbf{c}^{(T)}_{t}+\mathbf{b}_{\mathrm{out}}
\in\mathbb{R}^{H}
\label{eq:residual_head}
\end{equation}

The teacher model is trained offline using a trajectory-level squared-error loss over the next $H$ future steps:
\begin{equation}
\mathcal{L}_T
=\frac{1}{H}\sum_{h=1}^{H}\left\|
\left(P_t+\Delta\hat{P}^{(T)}_{t+h}\right)-P_{t+h}
\right\|_2^{2}
\label{eq:teacher_loss}
\end{equation}

This multi-step supervised learning encourages the teacher model to learn trajectory-level temporal patterns that are relevant to near-term ramps and bursty fluctuations.

\textbf{Point-wise student model.}
As shown in Fig.~\ref{Fig.2}, the proposed student model integrates a residual prediction head and an embedding head for feature-level knowledge transfer into a multilayer perceptron (MLP) backbone.
For deployment, we construct a compact MLP-based student model that performs constant-time inference using only point-wise input $(P_t,\mathbf{x}_t)$.
The proposed student model adopts a deliberately simpler architecture than the teacher model, because it is intended for deployment as a low-overhead rolling predictor rather than as a high-capacity sequence model.
Let $\mathbf{v}_t=[P_t;\mathbf{x}_t]\in\mathbb{R}^{d_u}$ denote the point-wise input vector.
A two-layer MLP is used to produce hidden representations:
\begin{equation}
\mathbf{h}^{(S)}_t=\sigma(\mathbf{W}_{s1}\mathbf{v}_t+\mathbf{b}_{s1})
\label{eq:student_h}
\end{equation}
\begin{equation}
\mathbf{r}^{(S)}_t=\sigma(\mathbf{W}_{s2}\mathbf{h}^{(S)}_t+\mathbf{b}_{s2})
\label{eq:student_r}
\end{equation}
where $\mathbf{W}_{s\cdot}$ and $\mathbf{b}_{s\cdot}$ denote the learnable weight matrices and bias vectors of the first and second layers. Two task heads are applied on $\mathbf{r}^{(S)}_t$ to produce the one-step residual prediction and the student embedding for feature distillation. The first head outputs the one-step residual prediction:
\begin{equation}
\Delta\hat{P}^{(S)}_{t+1}=\mathbf{w}^{\top}_y\mathbf{r}^{(S)}_t+b_y
\label{eq:student_residual}
\end{equation}

The second head produces a student embedding for feature distillation:
\begin{equation}
\mathbf{z}^{(S)}_t=\mathbf{W}_z\mathbf{r}^{(S)}_t+\mathbf{b}_z
\label{eq:student_embed}
\end{equation}

The final one-step absolute forecast of the load consumption is obtained by
\begin{equation}
\hat{P}^{(S)}_{t+1}=P_t+\Delta\hat{P}^{(S)}_{t+1}
\end{equation}

The proposed student model produces a one-step forecast at each control interval and can be recursively invoked for predicting a multi-step future trajectory.
This compact point-wise design is adopted to reduce online inference overhead under deployment constraints.

\subsection{Distillation through Sequence-to-Point Transfer}

As shown in Fig.~\ref{Fig.2}, a distillation algorithm is developed to transfer trajectory-level predictive knowledge from the offline teacher to the compact point-wise student model.
To update only the student parameters based on the well-trained teacher model, logit distillation for short-horizon behavior alignment and feature distillation for representation transfer are proposed. Such knowledge distillation effectively preserves the teacher model's most useful temporal knowledge in a form compatible with point-wise online inference.

\textbf{(i) Sequence-to-point logit distillation.}
Directly matching the student's one-step output to the teacher's full $H$-step trajectory is structurally incompatible and may unnecessarily over-constrain the deployed predictor.
Instead, the teacher's trajectory-level predictive behavior is compressed into a scalar soft target that remains consistent with one-step deployment while still reflecting the teacher's short-horizon predictive behavior.

Let $\hat{\mathbf{P}}^{(T)}_{t+1:t+H}$ denote the teacher model's $H$-step forecast. A convex projection operator is used to convert the teacher's predicted trajectory into a single soft supervisory target while preserving the physical scale of the forecast:
\begin{equation}
\tilde{P}^{(T)}_{t+1}
=\Pi\!\left(\hat{\mathbf{P}}^{(T)}_{t+1:t+H}\right)
=\sum_{h=1}^{H} w_h\,\hat{P}^{(T)}_{t+h}
\label{eq:convex_projection}
\end{equation}\\
where $\sum_{h=1}^{H} w_h=1$ and $w_h\ge 0$.
The weights $w_h$ can be chosen to emphasize the earliest horizon steps, for example, through a decaying pattern, so that the projected target places greater importance on imminent dynamics most relevant to real-time coordination, such as ramp-limit risks.

The resulting logit distillation loss is defined as
\begin{equation}
\mathcal{L}_{\mathrm{logit}}
=\left\|\hat{P}^{(S)}_{t+1}-\tilde{P}^{(T)}_{t+1}\right\|_2^{2}
\label{eq:logit_loss}
\end{equation}
which encourages the student to align with the teacher not only at the hard-label level, but also at the level of softened short-horizon predictive behavior.

\textbf{(ii) Feature distillation.}
To transfer the latent temporal representations learned by the teacher model, we further distill the temporally aggregated context vector $\mathbf{c}^{(T)}_t$, which is obtained by the attention pooling over the teacher's encoder output in \eqref{eq:context_vector}.
Since the teacher context vector and the student embedding may have different dimensions, 
the teacher context vector $\mathbf{c}^{(T)}_t$ is projected into the student embedding space:
\begin{equation}
\mathbf{z}^{(T)}_t=\phi_T\!\left(\mathbf{c}^{(T)}_t\right)
=\mathbf{W}_t\mathbf{c}^{(T)}_t+\mathbf{b}_t
\label{eq:teacher_proj}
\end{equation}
where $\mathbf{z}^{(T)}_t$ denotes the projected teacher embedding used as the feature-level distillation target.

This feature-level alignment, acting as a sequence-to-point knowledge transfer, enables the deployed student model, which only receives point-wise input, to approximate a compact representation induced by the teacher's temporally aggregated sequence understanding. The loss function for the feature distillation in \eqref{eq:teacher_proj} is calculated to minimize the discrepancy between the student embedding and the projected teacher embedding, as
\begin{equation}
\mathcal{L}_{\mathrm{feat}}
=\left\|\mathbf{z}^{(S)}_t-\mathbf{z}^{(T)}_t\right\|_2^{2}
\label{eq:feat_loss}
\end{equation}
where the student embedding $\mathbf{z}^{(S)}_t$ is attained by \eqref{eq:student_embed}. As a result, the proposed student model can learn a compact representation aligned with the teacher’s temporally aggregated context.

To this end, the proposed knowledge-distilled student model is trained with three complementary objectives: ground-truth supervision via minimum square error (MSE), logit distillation for short-horizon behavior alignment, and feature distillation for representation transfer. The composite loss function of the proposed student model is calculated by combining \eqref{eq:logit_loss} and \eqref{eq:feat_loss} with the MSE term:
\begin{equation}
\mathcal{L}_S
=\mathcal{L}_{\mathrm{mse}}
+\mathcal{L}_{\mathrm{logit}}
+\lambda\mathcal{L}_{\mathrm{feat}}
\label{eq:student_objective}
\end{equation}
where $\mathcal{L}_{\mathrm{mse}}
=\left\|\hat{P}^{(S)}_{t+1}-P_{t+1}\right\|_2^{2}$ keeps the student model directly anchored to the observed forecasting target and ensures that the deployment model remains optimized for actual next-step prediction accuracy rather than merely imitating the teacher;
$\lambda \in [0,1]$ is a trade-off coefficient controlling the strength of feature-level transfer.
The coefficients of $\mathcal{L}_{\mathrm{mse}}$ and $\mathcal{L}_{\mathrm{logit}}$ are fixed to 1 so that student training is primarily driven by fidelity to the observed next-step target and alignment with the teacher's softened short-horizon predictive behavior.
The feature distillation serves as an auxiliary representation constraint that enriches the deployment model without overwhelming the primary forecasting objective.

\section{Case Study}
The proposed deployment-efficient AI data center load forecasting algorithm is tested on the MIT Supercloud dataset \cite{Samsi2021}. The raw dataset is collected from a high-performance computing (HPC) system and contains $\sim 2\,\mathrm{TB}$ of time-series telemetry from GPU-accelerated nodes.
After resampling to 1-min resolution, multi-channel computing-related features are used, including GPU and memory utilization and temperature, per-node power (summed from per-GPU power), and multiple workload indicators such as job count, job switch, and GPU count.

All variables are min--max normalized based on the training split only.
The data is divided chronologically into 70\% training, 15\% validation, and 15\% testing.
This experimental setting aligns with an edge computing deployment scenario at control-layer nodes, where forecasting must satisfy tight latency and memory budgets.

Model performance is evaluated in terms of both forecasting accuracy and deployment efficiency. To quantify forecasting accuracy, mean absolute errors (MAEs) and root mean squared errors (RMSEs) are computed on the normalized targets over the chronologically held-out test set for short-term load forecasting. To measure deployment efficiency for lightweight inference, four metrics are adopted, including (i) the number of trainable parameters, (ii) on-disk model size, (iii) parameter memory in single-precision 32-bit floating point (FP32), and (iv) CPU inference latency (mean/p95) under a fixed batch size.
All experiments are conducted on a workstation with an Intel Xeon E5-2680 v4 CPU (2.4\,GHz) and an NVIDIA GeForce RTX 3080~Ti GPU.

A hybrid strategy combining manual tuning and grid search is adopted for the model fine-tuning. Both the teacher and student models are optimized using AdamW with an initial learning rate of $1\times10^{-4}$. The encoder-only attention-based teacher model uses two encoder layers with model dimension $d_m=64$, $K=4$ attention heads, and a feed-forward hidden dimension of 128. The forecasting settings are fixed to a historical window of $L=360$ and a prediction horizon of $H=15$. The lightweight student model is a two-layer MLP with hidden size 64 and a 16-dimensional projection head for feature distillation. The feature distillation weight is set to $\lambda=0.1$, from a grid search from $\{0.05,\,0.1,\,0.2\}$. 

\subsection{Forecasting Performance Comparison}
 We compare the forecasting accuracy and model size of the proposed method with those of two recent deep learning algorithms, namely GRU- and LSTM-based models. The proposed method is implemented with residual learning, while the GRU- and LSTM-based baselines follow the vanilla design.

\begin{figure}[!t]
  \centering
  \subfloat[]{
    \includegraphics[width=0.42\textwidth]{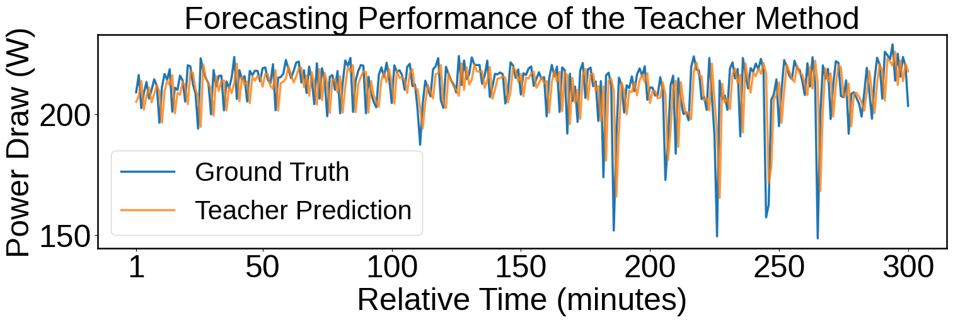}
    \label{fig:teacher_vis}  }\\
  \subfloat[]{
    \includegraphics[width=0.42\textwidth]{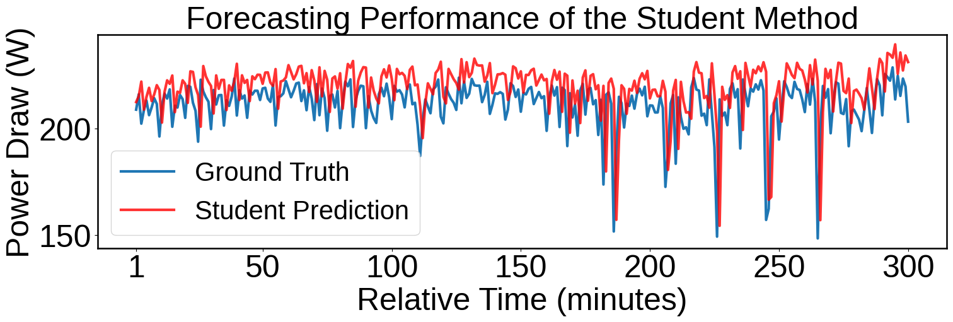}
    \label{fig:student_vis} }
  \caption{Comparison of forecasting performance with the ground-truth load values. (a) Transformer-based teacher model through $H$-step trajectory forecasting; (b) Lightweight knowledge-distilled student model through one-step recursive rolling forecasting.}
  \label{Fig.3}
\end{figure}

Fig.~\ref{Fig.3} illustrates the predicted trajectories by the proposed high-capacity teacher and lightweight student models over the first 300 test time steps. It can be observed that the teacher model accurately captures the load consumption trajectory and successfully tracks the dynamics induced by abrupt job switches. Importantly,  the distilled student model by the proposed method accurately tracks the major burst onsets, local ramps, and short-term fluctuations with high fidelity. Although the student model is slightly less accurate than the Transformer-based teacher model, it preserves the overall bursty dynamics learned through the sequence-to-point distillation while enabling low-latency rolling inference for lightweight deployment. 

\begin{figure}[!t]
  \centering
    \includegraphics[width=0.495\textwidth]{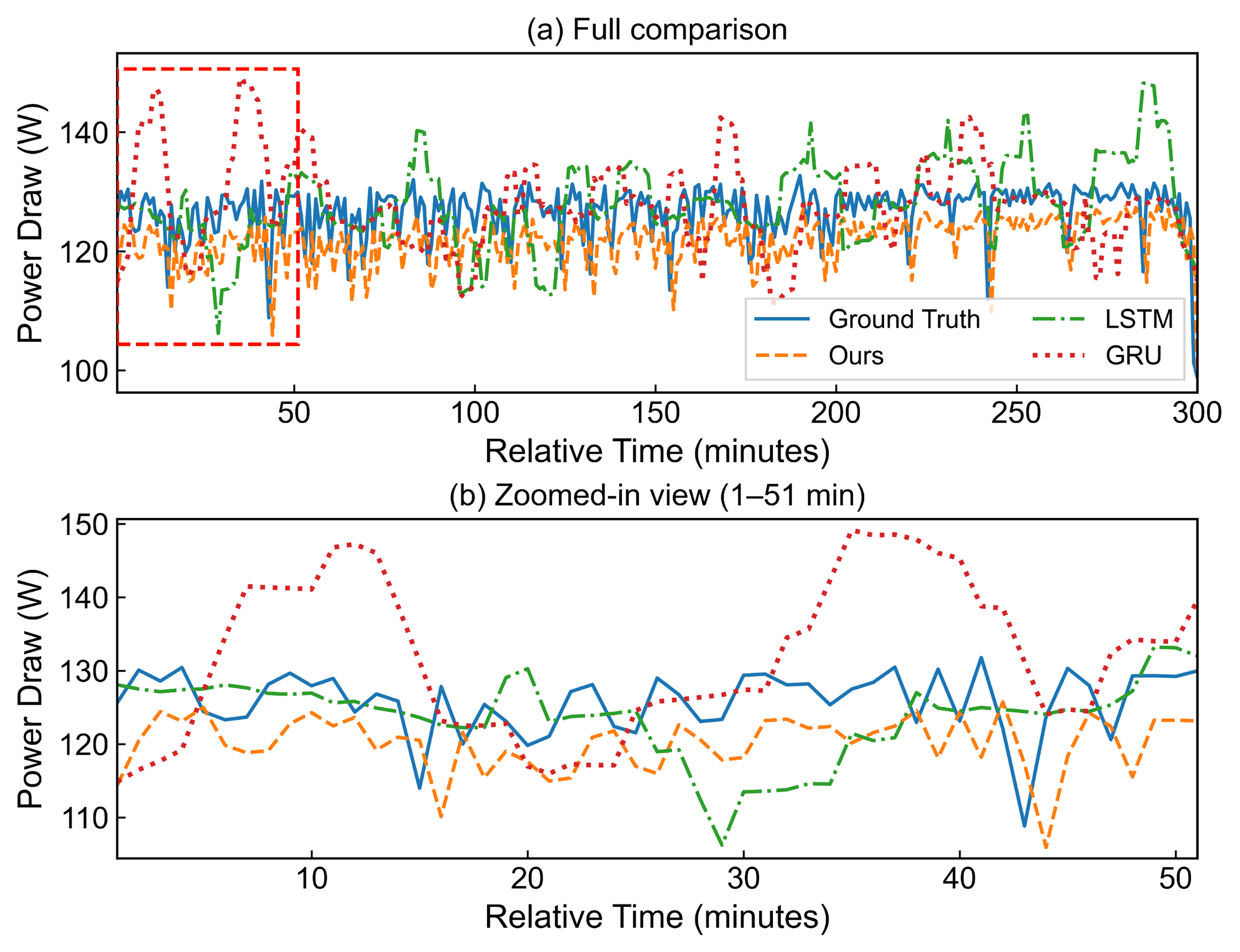} 
  \caption{Forecasting performance comparison of the LSTM-based, GRU-based, and the proposed student model.}
   \label{fig:comparison}
\end{figure}

Furthermore, Fig.~\ref{fig:comparison} compares the predicted trajectories by the LSTM-based model, the GRU-based model, and the proposed distilled student model at a 1-minute resolution, with a zoomed-in view provided for clearer visual comparison. It can be observed that the proposed student model tracks the ground-truth trajectory more closely than the compared baselines over the 300-minute time window. Table~\ref{tab:accuracy} compares the MAEs and RMSEs of these agorithms over all the test instances under one-step-ahead forecasting. As shown in Table~\ref{tab:accuracy}, the proposed teacher model achieves the highest forecasting accuracy among the compared methods, yielding the lowest MAE of 3.14\% and RMSE of 9.96\%. 

\textbf{It should be noted that the proposed student model achieves forecasting accuracy comparable to that of the teacher model, with an MAE of 3.25\%.} Both proposed teacher and knowledge-distilled student models outperform the LSTM- and GRU-based baselines. Among the benchmark methods, LSTM achieves the best performance for AI data center load forecasting, as reported in \cite{mughees2025short}. Compared with the LSTM, the proposed student model reduces MAE by 1.87 percentage points and RMSE by 3.46 percentage points, which correspond to improvements of 36.5\% and 25.4\%, respectively. Relative to GRU, the reductions are 2.10 percentage points in MAE and 3.91 percentage points in RMSE.

Beyond forecasting accuracy, Table~\ref{tab:deployment} compares deployment-centric metrics, including the number of trainable parameters, FP32 parameter memory, on-disk model size, and CPU inference latency. It is shown that the proposed student model substantially shrinks the deployment footprint compared to other deep neural models. \textbf{Compared with the state-of-the-art LSTM-based method, the proposed student model reduces the number of trainable parameters from 81.9K to 6.1K, corresponding to a 13.4× reduction}. Meanwhile, such a lightweight model reduces FP32 parameter memory from 320.00 KB to 23.82 KB, on-disk model size from 520.00 KB to 27.74 KB, and the CPU latency from 4.8 ms and 6.2 ms to 0.6 ms and 0.9 ms for mean and p95 latency, respectively. The GRU-based method has 64.0k parameters, and our method has only 6.1k, achieving approximately 10.5 times fewer parameters. Similar reductions are observed in FP32 memory, on-disk size, and CPU latency. To this end, the proposed student model is markedly more efficient in deployment, with substantially lower parameter count, memory usage, storage size, and inference latency. These results show that the proposed sequence-to-point distillation algorithm effectively balances forecasting accuracy and model deployment.
This allows the student model to preserve much of the teacher model’s temporal modeling capability while remaining lightweight enough for real-time rolling inference.

\begin{table}[!t]
\centering
\captionsetup{font=small,justification=centering}
\caption{\textsc{Comparison of Forecasting Accuracy}}
\label{tab:accuracy}
\renewcommand{\arraystretch}{1}
\begin{tabular}{|>{\centering\arraybackslash}m{0.35\columnwidth}|
                >{\centering\arraybackslash}m{0.20\columnwidth}|
                >{\centering\arraybackslash}m{0.20\columnwidth}|}
\hline
\multirow{2}{*}{Compared Model} & \multicolumn{2}{c|}{Prediction Errors} \\ \cline{2-3}
& MAE [\%] & RMSE [\%] \\ \hline
GRU    & 5.35 & 14.09 \\ \hline
LSTM   & 5.12 & 13.64 \\ \hline
Proposed student model & 3.25 & 10.18 \\ \hline
\textbf{Proposed teacher model} & \textbf{3.14} & \textbf{9.96} \\ \hline
\end{tabular}
\end{table}

\begin{table}[!t]
\centering
\vspace{8pt}
\captionsetup{font=small,justification=centering}
\caption{\textsc{Comparison of Deployment-Centric Metrics}}
\label{tab:deployment}
\renewcommand{\arraystretch}{1}
\begin{tabular}{|>{\centering\arraybackslash}m{0.21\columnwidth}|
                >{\centering\arraybackslash}m{0.1\columnwidth}|
                >{\centering\arraybackslash}m{0.15\columnwidth}|
                >{\centering\arraybackslash}m{0.13\columnwidth}|
                >{\centering\arraybackslash}m{0.16\columnwidth}|}
\hline
\multirow[c]{2}{*}[-2.5ex]{\#Model} & \multicolumn{4}{c|}{Deployment-Centric Metrics} \\ \cline{2-5}
& Params. & FP32 Mem. [KB] & On-disk [KB] & CPU Latency [ms] \\ \hline
GRU    & 64.0K & 250.00 & 430.00 & 4.1/5.3 \\ \hline
LSTM  & 81.9K & 320.00 & 520.00 & 4.8/6.2 \\ \hline
\textbf{Our student} & \textbf{6.1K} & \textbf{23.82} & \textbf{27.74} & \textbf{0.6/0.9} \\ \hline
Our teacher & 69K & 270.00 & 410.00 & 8.5/11.0 \\ \hline
\end{tabular}
\end{table}

\subsection{Value of Proposed Distillation}

To assess the contribution of the distillation, we conduct an ablation study while keeping the student backbone and training protocol unchanged. 

As shown in Table~\ref{Table 2}, the student model trained without distillation yields the worst performance, indicating that ground-truth supervision alone is insufficient for a compact point-wise predictor to capture bursty short-term load dynamics. 
Adding logit distillation reduces the MAE from 4.15\% to 3.25\% and the RMSE from 11.80\% to 10.18\%, suggesting that alignment with the teacher's softened short-horizon forecast effectively transfers behavior-level knowledge. 

The proposed logit and feature distillation jointly yield an MAE of 3.25\% and an RMSE of 10.18\%. This is because the logit distillation in the proposed method improves short-horizon output alignment, whereas the feature distillation enhances the student model's internal representation of sequence-aware temporal context.

\begin{table}[!t]
\centering
\captionsetup{font=small, ,justification=centering}
\caption{\textsc{Comparison of One-Step Forecasting w and w/o Distillation}}
\label{Table 2}
\renewcommand{\arraystretch}{1}
\begin{tabular}{|>{\centering\arraybackslash}m{0.4\columnwidth}|
                >{\centering\arraybackslash}m{0.2\columnwidth}|
                >{\centering\arraybackslash}m{0.2\columnwidth}|}
\hline
\multirow{2}{*}{Compared Model} & \multicolumn{2}{c|}{Prediction Errors} \\ \cline{2-3}
& MAE [\%] & RMSE [\%] \\ \hline
Student w/o distillation       & 4.15 & 11.80 \\ \hline
\textbf{Proposed student model} & \textbf{3.25} & \textbf{10.18} \\ \hline
\end{tabular}
\end{table}

\section{Conclusion}
This paper proposes a deployment-efficient knowledge distillation-based short-term load forecasting method in AI data centers. Case studies demonstrate that the approach captures bursty, non-stationary, workload-driven dynamics via an offline sequence teacher model while enabling low-latency rolling inference through a lightweight pointwise student for edge deployment. By joint logit and feature distillation, the proposed sequence-to-point knowledge transfer algorithm achieves competitive forecasting accuracy while substantially reducing the deployment footprint. 
Furthermore, the resulting lightweight forecasting method provides a novel deployment-efficient building block for the subsequent flexibility management and grid-data center coordination. It has the potential to enable flexible job switching and smooth under bursty AI workloads for demand response, which will be our future work.

\IEEEpubidadjcol

\bibliographystyle{IEEEtran}
\bibliography{IEEEabrv,Citation}
\let\mybibitem\bibitem

\end{document}